# Reconstructing Bologna
## The City as an Emergent Computational System – A Study in the Complexity of Urban Structures

## Part I: The Basic Idea & Fundamental Concepts

Rainer E. Zimmermann,

IAG Philosophische Grundlagenprobleme,
FB 1, UGH, Nora-Platiel-Str.1, D – 34127 Kassel /
Clare Hall, UK – Cambridge CB3 9AL[1] /
Lehrgebiet Philosophie, FB 13 AW, FH,
Lothstr. 34, D – 80335 Muenchen[2]
e-mail: pd00108@mail.lrz-muenchen.de

Anna Soci*, and Giorgio Colacchio

Dipartimento di Scienze Economiche,
Facoltà di Science Politiche,
Università degli Studi,
Strada Maggiore 45, I – 40125 Bologna /
*Clare Hall, UK – Cambridge CB3 9AL
e-mail: soci@spbo.unibo.it, si10037@iperbole.bologna.it

## Abstract

The conceptual background for a detailed study of the urban form of the city of Bologna is discussed with a view to modern methodological insight as it is being presented by recent results of complexity theory and the theory of self-organized criticality. The basic idea is to visualize the city of Bologna as an example of a massively parallely organized and interacting complex computational system in the sense of these recent theories. It is proposed to relate aspects of urban evolution to a universal concept of evolution which is governing all processes in nature. The universality of this approach is thought of as being an epistemological advantage as compared to more classical studies utilizing primarily local and specific methods for their modelling procedures. To actually establish whether this is in fact an advantage or not will be one of the main results of this present series of papers. In this very first part of the study, the basic idea is explicated in some detail, and the fundamental concepts are introduced in order to clarify the terminology utilized in the following. Two more parts of this study will follow in due time. One will discuss the empirical data actually available [1], the other will discuss more general consequences of this approach. [2] Some basic aspects of the more technical viewpoint of the methods put forward here have been discussed in two preliminary papers [3],[4], stressing the underlying motivation which points towards a general concept of self-organizing emergent systems.

---

[1] Permanent Addresses.
[2] Present Address.

## Introduction

As to the explicit construction of its urban form, Bologna has had a revolutionary past. In particular, with respect to its historical centre, the structural development of a progressive conservation of historical urban substance has been prepared in an exemplary (if not paradigmatic) manner. The original programme of sanitation and rehabilitation ( = re-construction) and its mediating presentation (Piano per il centro storico, 21$^{st}$ July 1969, cf. Pier Luigi Cervellati: Commune di Bologna, Centro storico, 1970), both the outcome of a period of scientific coordination by the study group of Leonardo Benevolo between 1962 and 1965, have been milestones of modern city planning. For the first time, as far as we can recognize today, the methodological conception of a scientific as well as historical and morphological urban analysis has been laid down with a view to treating the historical centre as a complete monument in its totality. [5] A number of innovative concepts have been decisive for the practical application of this approach in daily urban life for a long time.
First of all, it is the principle of an explicitly *de-centralized public organization* which has been the main pillar of this urban structural architecture: The idea was to take *quarter councils* and *quarter assemblies*, respectively, as means of a democratic integration of decision making by transparent, self-organizing groups as counterparts of the private sphere of the family organization (decentramento). An immediate corollary of this approach was the necessity of establishing an urban structure of *nearness and vicinity* aiming towards the concrete public life *in the place* rather than of drafting an abstract scheme of transitory traffic flows. As documented later at the first world conference of urban traffic (10-12 June 1974) when the „declaration of Bologna" was signed, mobility is visualized within the framework of this conception as one of the most essential necessities of human life. Its fulfilment should be satisfied therefore such that security, comfort, convenience, and swiftness can be actually harmonized. In this sense, transport is being defined as a *public service* which must not be organized according to principles of profit, but only according to its social utility. In particular, the rights of the pedestrian should be visualized as part of the human (declaration of) rights. (cf. [6], pp. 103sq.) In the sense of this declaration, the administration of Bologna had already introduced a *utility tariff* of public transport with a zero rate for all during rush hour traffic, and free of charge for pupils and students during school hours and university terms, and for old age pensioners alto-gether. The main underlying objectives of this approach are still today part of the urban planning objectives in so far as they outline the basic achievement of „a sustainable mobility through a decrease in energy consumption, a reduction of pollution, the offer of adequate accessibility [as well as] the recovery of urban areas unduly invaded by cars." ( cf. [7], p.1)

Obviously, these basic measures taken, although of considerable progressivity at the time of their initiation in the middle of the sixties of the last century, well before ecological aspects belonged to the standard inventory of urban planning, would not have been sufficient in order to actually change the traffic situation in the city which at the time was not so different from other Italian (and probably European) cities. Hence, a catalogue of supporting measures was introduced: *Parking space* was reduced and de-centralized at the city edge, with a view to a concept which later on became known in other places as „park & drive". Parking houses and other parking space attractors due to large building complexes or shopping centres were being banned from the city centre. The *unification* of the various public bus services, mainly a result of the conception put forward by Mauro Formaglini in June 1972, and the decisive enlargement of the *car park* (from 285 buses to 513 between 1972 and 1975) helped to substitute private motor car traffic by public bus services. (cf. [6], pp. 85, 93sqq.) Finally, a number of side-measures supported this approach by introducing new *pedestrian zones*, not only within the city centre, by *reserving certain streets* for residents, supply roads for firms, tracks and lanes for buses and taxis, connective roads for exclusive use at certain times of the day, and so forth. In 1974, a reduction of unregulated street length had been achieved to 140 out of a total of 580 km. (cf. [6], p. 96) The whole conception made it possible to moderate the urban dimensions such as to fit them to pedestrian needs, in order to orient oneself within and to traverse quickly and conveniently through the available network of bifurcating streets, lanes, squares and places, with short cuts among *palazzi* and guarded by the traditional *portici* characteristic for the city structure, partitioning the latter into typical relative distances such as lane width, distances between neighbouring buildings, and so forth. The actual hierarchy introduced into the system of street connections (*rete primaria* and *rete secondaria*), regulated in terms of explicit signalling systems indicating turning prohibitions or commandments, one way streets, and diversions, and in terms of extension rules (such as demanding 30-50 m$^2$ of „green" land for each 100 m$^2$ of new parking space at the edge), had obvious implications for the *settlement of shopping centres* and other commercial institutions: The idea was to let private shopping co-operatives, grown market places, and „flying trades" institutions prevail and block the city instead against the inflow of large commercial centres. The *Centro Marco Polo* in the *Lame* quarter (4$^{th}$ district) was one of the first co-operatives settling at the habitat construction of Beverara (Piazza Giovanni da Verrazzano) opened in 1974.

It is not the appropriate place here to discuss the actual difficulties encountered by the city administration as presented by the Italian national state economy, and in particular by national laws and decrees issued by the central government aiming at bringing the local government back to the „general track" again. This is especially true for a decree dealing with an explicit definition of consummation regulations as to the settlement of various shopping institutions within the city centre, which in the end led to an adminstrative strategy of amending an actual law (no. 426, art. 9) by a „footnote" exempting it for the whole of the city. See [6] for further details. But it should be noted that the directional relationship between local province and central government in Italy is alien to those who live in a federation of countries or states with autonomy of the latter in certain well-defined fields such as education, local trade regulation, public transport, and so forth. Obviously, this is a framework of global boundary conditions to any processes initiated on the local level which has to be taken into account when discussing the political and ethical implications of the approach we are dealing with here.

But note also something else: As we can clearly recognize, the formal (re-) construction of the urban structures involved displays manifold connections with the explicit processes of everyday life in a community. In fact, as it turns out, the results exhibited by a close analysis of the various aspects of city life point towards what we may justifiably call „metaphysical implications". This viewpoint is still not very common within European science. On the other hand, nobody would doubt the justification of Asian „metaphysical" architecture when applying principles such as those of the *Feng Shui* to the actual erection of buildings. (In fact, the Needham institute in Cambridge (UK) is built according to these principles.) And it is not very difficult to recognize harmonizing aspects within Asian city structure such as displaying the existence of „difference within a framework of repetition". However, as Henri Lefebvre has pointed out some time ago ( cf. [8], pp. 157sq.), this Asian structure of social space has one serious drawback: it is an *attribute of power*. It implies and is implied by divinity and empire, and it thus combines knowledge with power. Nature (whether available or constructed) is harmonized according to a hierarchy of principles which explicitly centres around the emperor, i.e. the representation of a feudal system. Obviously, this is something which explicitly contradicts the insight gained in the period of European enlightenment. Hence, contrary to the Asian approach, introducing the *concept of de-centralization* into the city structure reflects a fundamentally European approach so as to deal with these „metaphysical" aspects of daily life. Indeed, as we will find out later, de-centralization is not only a secularized principle of social organization, but also a basic principle of evolutionary processes abundant in nature (if visualized in terms of modern European science). This illustrates clearly that ethical implications of this „metaphysical" background are always present: In fact, we realize this in the very catalogue of (technical) measures taken when thinking of the underlying objectives showing up within the details of urban planning. It is not a coincidence that these implications are tied to legal aspects of democratic policies such as defining technical rights (e.g. of pedestrians) with respect to the human rights convention. „Quality of life" is a pertinent concept showing up in the motivation of arbitrary measures eventually taken – so as to introduce the elimination of scriptures on the walls of historical buildings with the words: „Al fine di migliorare la qualità di vita dei cittadini ..." [„Finally, in order to improve the quality of life for the citizens ..."](cf. [9], p.2)

However, before we can establish this insight as a result of our discussion, we have to introduce the fundamental terminology and the concepts according to which we would like to proceed in the following. Hence, in section 1, we will discuss the concepts of complexity and emergence in all generality. In section 2, we will define computational systems and display their relationship to recent results of the theory of self-organized criticality in the sense of the Santa Fe school. The context dependence of various aspects of this theory will be important for its application to social systems. This will be the topic of section 3. The particular relationship of urban structures to social space will be discussed in section 4. In section 5, we will show that the actual choice taken in urban planning represents underlying ethical consequences which are part of the planning process in the first place. The final section 6 will serve to outline the further programme of this present study dealing with the re-construction of the historical centre of Bologna. A number of appendices is given in order to point to side aspects of this research.

# 1   Complexity & Emergence

As Edmonds has shown in his recent PhD thesis ([10], p.19), the problem of actually determining the meaning of emergence and the various levels of complexity accessible is deeply related to the modelling procedure itself: Hence, one of the objectives of this present series of papers shall also be to demonstrate that it is possible, in principle, to eventually develop models which are more than vaguely connected to models in different fields, very much on the line of the overall objective of this project. In fact, as it turns out, the models utilized here, may well satisfy both the definitions for being syntactic and semantic, at the same time. ([10], pp.25sq.) It may be noticed however, that we would not tend to underestimate the impact of *metaphorization* within the explicit modelling procedures, as is perhaps done sometimes by Edmonds when referring to Braithwaite. ([10], p.28) Contrary to that, we would stress the point that metaphorization will play an important role, and it is *analogy* then which is lying at its root. Moreover, we will also utilize both the *predictive* and *explanatory* modes of models for our purposes. This is so because (consistent with what has been said above) we aim at the explanation of the phenomena involved in more general evolutive (and thus universalized) terms, but *also* at the developmental prognosis of the characteristic parameters actually defining the models. While generalizing the argument of Edmonds somewhat, we agree that indeed, the difficulty of finding such a configuration of models – namely one which is able to closely approximating the actual data while also explaining what happens – can be used as a *measure for the complexity* of the system which is to be investigated. ([10], pp.29sqq., 39) Continuous modelling can serve as a first approach towards clarifying the processes involved in constituting such a model configuration. (Cf. Appendix 1) On the other hand, there is a number of aspects to that which deserve closer inspection: the relationship among the concepts of emergence, complexity, and contextuality, as well as the relationship among all of those and the concepts of computation and information.

As to emergence, the first point is that of *innovation*. Crutchfield has shown that the actual difficulty is that of discovering anything new at all when what we can describe can only be expressed in terms of a language which reflects the current understanding. ([15], p.1) This problem has also been addressed in some detail when dealing with the Paris scenario. [11] As it turns out, the point is that „newness is in the eye of the observer" ([15], p.3) – very much in the same sense, as complexity is, visualized as something which is a *property of the description* of a system rather than a property of the system itself. ([16], p.2) With what we deal then is languages rather than anything else: „An emergent phenomenon is one that is described by atomic concepts available in the macro-language, but cannot be so described in the micro-language." ([16], p.4) We will come back to this in section 3.

For complexity then, it follows that it is the interactive organization of systems which when being mapped onto a model turns out to be a mixture of order and disorder as it is being observed by the one who models. In this sense, we can recognize that on the one hand it would be appropriate to classify the complexity of models in the first place by means of some combinatorial characteristic of the modelling procedure. This is actually being done by Edmonds who defines complexity as „that property of a model which makes it difficult to formulate its overall behaviour in a given language, even when given reasonably complete information about its atomic components and their interrelations." ([10], p.72 – see also the qualifications made there with respect to the references) If the model can be represented in terms of a suitable graph with a well-defined tree structure, then the combinatorial characteristic being used in order to classify the model's complexity is given as the *cyclomatic number* of the respective graph (which is essentially the number of connecting arcs minus the number of nodes plus one). ([10], p.108) The hierarchy of the diagram nodes exhibits nothing but a syntactic structure. And it is this structure which decides in the end about which parts of a modelled system are taken to be „atomic", thus determining the micro-behaviour of a system rather than its macro-behaviour.

So far the syntactic side. It is contextuality then, which eventually determines the semantic side. Crutchfield points out that a given state of a system has to be visualized with respect to the set of future sequences that could be observed (called *morphs* of this state) essentially constituting the *context* in which some observation and/or measurement take on semantic content. ([15], p.19) This demonstrates the (pragmatic) necessity of introducing contextuality into the con-cept of complexity at all which is not common yet for the most of science. ([16], pp.1sq.) But this approach is well consistent with the common sense approach of visualizing complexity as an instrument of comparison rather than an instrument of measurement, in the first place. ([10], p.46) The better then, if the utilization of graph theory enables us to introduce a combinatorial measure of some kind. In the end however, this points strongly to context-dependent *information* altogether. It appears as if contextuality would actually *qualify* information.

It is important to notice here that the model complexity referred to in this present paper is not identical with the complexity of information. Instead, the latter appears to be an approximation of the former. This can be seen in the following way: The basic idea of perception and cognitive grasping of the world is that one needs information storage. Stability and order are then necessary conditions for such a storage. However, instability and disorder are necessary conditions for actually *producing* new information and for communicating it. The trade-off between these two tendencies is what probably computation theory is about, as Crutchfield has argued earlier. ([15], p.46) But information, in so far as its complexity is actually being measured, is quantized according to „size" (size as program length, size as number of elements of a set of rules). And although „size" may turn out to be a necessary condition for complexity, it does not appear to be a sufficient condition, because, intuitively, one can imagine large systems which are quite simple. In this sense, algorithmic information complexity as introduced by Kolmogorov is an appropriate measure for information, but not for complexity. In other words: The quantity of information shows up as a weak approximation to a model of complexity. ([10], pp.57sqq.)

Finally then, what is the state of computational systems within this framework of complexity measures? In principle, a process in nature can be said to be understood, if an adequate model of it can be re-constructed. Computation shows up within this framework as a concrete mapping of the (moving) process structure. Hence, using a minimal model with a minimal error, then the optimal predictor of the process being able to reproduce the actual measurements can be visualized as a machine whose structure is a suitable approximation to the process. This machine however, is based on the information processing component of this modelling alone. In particular, the machine's architecture represents the actual organization of this information processing. And this is nothing but computation. Taking biological evolution as an example, we realize that in order to properly survive, an agent has to model his environment adequately. This can be achieved only with respect to the available computational resources and the explicit language utilized. Innovative understanding is then the capability to perform a transition from one such model language to another such that the old one is a special case of the new one. (Cf. [15], pp.8sqq.) For this (restricted) perspective alone are we able to talk about „computational complexity" which can be visualized as the evolution from disordered states which are practically incompressible programs. Co-evolutionary processes turn up then as being co-computational. [17] Hence, the performing of a computation corresponds to a temporal sequence of changes in some system's internal state (ascribed in terms of measurements made with respect to a given model of the system). In this case, Shannon's entropy rate can be defined in a suitable way as a measure of the rate with which the environment appears to produce information (in bits per symbol). The higher the rate, the more information produced, and the more unpredictable the environment. ([15], pp.11, 20) Visualized in computational terms then, a process can be said to undergo emergence, if the architecture of information processing changes spontaneously such that a distinct and more powerful level of computation shows up that was not present before. ([15], p.67) In terms of Shannon's entropy formula, complexity of some description can be given as

$$C(x) = \lim \lambda \log_2 N - \log_2 \omega(\lambda, x),$$

where C is measured in bits, $\lambda(x)$ is the length of the description, N the size of the alphabet used to encode the description, and $\omega$ the size of the class of all descriptions of length less than $\lambda$ equivalent to x. The limit is considered for $\lambda \to \infty$. For universal Turing machines, this is identical with the Kolmogorov complexity. ([16], p.2) But computational systems have also another aspect which will be discussed in the next section.

## 2   Computational Systems

The idea of this present work is to restrict urban complexity (and its theory) to the concept of visualizing a city as a self-organized emergent computational system as it is modelled by means of recently introduced methods originating mainly within the field of the natural sciences. The immediate starting point for this conception has been given by Crutchfield and Mitchell in motivating this evolutionary viewpoint. [18] For them, it is obvious that „[a]llowing global coordination to emerge from a decentralized collection of simple components has important advantages over explicit central control in both natural and human-constructed information-processing systems." ([18], p.1) This is mainly so, because centralized coordination implies heavy costs on its implementation: such as in terms of *speed* of the processing achieved (the organizing centre can turn out to be a bottleneck), *robustness* (breakdown of the organizing centre causes collapse of the network), and equitable *resource allocation* (the organizing centre acquires the largest share of resources for itself). Obviously, these aspects are of vital importance for the study of Bologna undertaken here, especially with a view to the explicitly *decentralizing organization* of the city put forward at the time of the reforms mentioned above. The relevant result of this paper of Crutchfield and Mitchell was that in a spatially distributed system which consists of many locally interacting processors, global computation may emerge as intrinsically strategic mode of this very system. In other words: Given a computational task which may be ascribed to a system altogether, specific computational strategies are being performed by locally interacting agents constituting the system such that eventually, adequate global results show up which are appropriate solutions of the original task. It is important to notice that this task cannot be covered by any individual strategy of any of the agents alone, but is to be considered as a „global" task of the system as a collection of these agents. This is exactly the situation we often encounter in economics: Competitive agents e.g. within a capital market, control their individual production investment and stock ownership strategies based on the optimal pricing that has emerged from their collective behaviour. ([15], p.4) To be more precise: These agents rate the optimal pricing with respect to the experience they gain locally. It is not necessary at all that they understand the global pattern of interactions and economic „laws" that govern the complete system. It is only essential for them that through the mediation of the market's collective behaviour, the prices which actually emerge turn out to be accurate signals which map the available information. What they do in practise is nothing but checking on the local consistency of the signals perceived. Hence, the idea is that (as visualized under the perspective of the global system) the system's computational capabilities are *themselves emergent*, in the sense that on the global level, additional functionality shows up as compared to all single local levels.

This implies that both levels are irreducible with respect to each other: The individual agent checks local consistency, and it is not necessary for him to actually know of and understand about the global system. On the other hand, knowledge of the global „laws" and properties of the system is not a sufficient condition for actually applying correct strategies on the local level. Basically, the chief problem for an agent is to predict the future sensory input in turn based on modelling the hidden environmental states and selecting possible actions. Obviously, this can be only done, if taking the local perspective. Global knowledge may help, but will not serve the purpose without local insight. In the case of visualizing the interactions within a city, this means that the complete city is being visualized as a global structure which has emerged as a result of superimposing local structures. These local structures are formed by the interactions among individual agents who do not necessarily know about all of the city. What they have to know however, is to rate consistently what they observe as sensory input from the immediate interactions within their own neighbourhood. In fact, they actually act *as if* they would automatically assume a self-consistency of the complete system between its local and global levels – without actually knowing that there *is* such a global level.

Crutchfield and Mitchell use a cellular automaton model for discussing their idea. This is something we will mention later again. (In fact, we will find out that urban modelling is also best visualized in terms of cellular automata, as compared to the more qualitative insight gained by means of continuous modelling as shown in the appendix 1.) The important point here is that the emergence of explicit strategies towards an increase of computational capacity points to a notion of „teleology" uncommon in science. If we think of a social system as one which is organizing itself with respect to its own computational capacities, then we have not to forget that this „self" is something we project onto the process due to our own linguistic condition, and that the process itself is nothing but a collection of response actions being undertaken under the impact of given boundary conditions. But these actions have their effectivity criterion in what we like to call „natural laws". The question is indeed whether social systems such as cities develop according to a lawful framework of strategy selection, if *all the other* systems in nature do so. To be more precise: whether our models of social systems would be compatible with other models exhibiting this characteristic. This is a problem which is closely related to another branch of Santa Fe studies, namely that of *self-organized criticality*. (Appendix 2)

The results as presented by Crutchfield and Mitchell were showing that evolution proceeds via a series of epochs connected by distinct computational innovations. The propagation of the structural information content was governed by the onset of *domains*, i.e. regions of recurring patterns. It was found out that regular domains of that sort were crucial for the organization of both the dynamical behaviour and the information processing properties of the underlying cellular automata structure. In particular, the propagation of *domain walls* (under spatial localization referred to as „particles") could be recognized as primary mechanism for carrying information over long distances. The emergence of domains, the propagation of domain walls, and the (particle) interaction among these, all of them turn out to represent the basic information used to define the system's intrinsic computation. As it appears, the localization and explicit determination of particle structures within such a computational system is the necessary first condition for eventually understanding the system's self-organizing computational behaviour.

On the other hand, we have to deal with the problem of „spaces of free play". The point here is the following: Per Bak, after having discussed several evolutionary processes in terms of directed percolation, in his book on self-organized criticality [19], notes that large fluctuations observed in (among others) economics indicate that an economy is operating at the self-organized critical state in which minor shocks can lead to avalanches of all sizes. There appears to be no way as to stabilize the economy with respect to these fluctuations. Hence, so he concludes, is there no possibility of actually influencing the processes so as to get rid of fluctuations which are rated in a negative manner and taken to be catastrophes of one or the other kind. ([19], pp.191sq.) In this sense, any technology designed to „smooth out" fluctuations would rather add to their increase. Bak visualizes therefore, the most robust state for an economy as one which is decentralized (!) and self-organized critical of capitalistic type with fluctuations of all sizes and durations. ([19], p.198) The point here is to ask whether humans would have any space of free play giving them the chance of somewhat manipulating instabilities so as to „improve" their states in economical (and other) terms. We would not only find the „agnostic" result of Bak's unsatisfactory with respect to the human acction potential discussed here, more than that, we would question that the appropriate type of economy must necessarily be of capitalistic type. (Not that we would, on the other hand, necessarily aim to rescuing Marx and/or Greenspan instead.) The question is rather whether agents which have their own computational capability of model building (as an actual outcome of the same process they are going to model) would not have appropriate means of taking their measures.

## 3   Context Dependence & Praxis

After what we have said so far, we realize now that what we aim at is to describe systems which are characterized by self-organized critical computational processes generating their own context. This is actually being achieved by a dynamical mediation of the micro-level and the macro-level of the respective system. To be more precise: In using an appropriate micro-language in order to describe the system's micro-level, the conditions are being produced for eventually attaining a macro-level of description. One of the first who discussed the practical consequences of such a procedure was Thomas Schelling. In his celebrated book of 1978 he laid down the basic aspects of this mediation of organizational levels in complex systems. [23] Concentrating on social systems, he explored the relation between the behaviour characteristics of the individuals who comprise some social aggregate, and the characteristics of this aggregate altogether. ([23], p.13) The crucial aspect is again the fact that „there is a notion of people having preferences, pursuing goals, minimizing effort or embarassement or maximizing view or comfort, seeking company or avoiding it, and otherwise behaving in a way that we might call ‚purposive'." ([23], p.17) Because all people do the same at the same time, all these purposes, objectives, goals and so forth are constrained by the environment such that their outcome is contingent. This is exactly the situation we encounter when exploring the urban behaviour of people within a spatial and temporal city structure. The important point is again that the individual needs to know a number of local details of the social interactions taking place, but does not need to know any global aspects of the complete system. ([23], pp.20, 22) Obviously, this relates to the ancient notion of Adam Smith who referred to the „invisible hand" bringing about the collective organization of all these individuals. Schelling's basic theme was to stress that it is possible to formulate propositions about the behaviour of the aggregate which are true globally, but not locally or in detail, and nevertheless, if true, independently so of how (individual) people behave. This discussion introduced not only the statistical law of large number, but also feedback loops and non-linear interactions. ([23], pp.49sq.) Again, one topic was contextuality (such as in the famous birthday example when the protagonist was regularly taking the kids to the Red Sox meeting always the same people in their seats vicinity) ([23], pp.40sq.), another the combinatorics of pairing from two populations (e.g. the „arithmetic" of marriage). In fact, it had been Lévi-Strauss somewhat earlier, together with André Weil, who had discussed the algebraic effects of certain types of marriage laws for the first time. [24] Schelling also discussed the consequences of the „commons problem" (the sharing of common ground in a community) [25] and of the percolation of income distributions.

Two main results were the outcome of this original book: the self-organization of communities in distributive patterns and the prisoner's dilemma. As to the first, Schelling could list a number of combinatorial criteria for the spontaneous onset of structures and patterns. As the most important turn out that great many phenomena actually occur in pairs due to dual complementary actions, that there are principles of conservation for closed systems as well as measurable quantities and countable densities for semi-closed systems, especially in terms of well-defined transition matrices, that there are variables which are rates of other variables, or independent variables which prove to be the sum (or additive superposition) of dependent variables, or the average of the behaviour that they induce, or different variables which have a common component. ([23], pp.76sqq.) Again, for the self-organized neighbourhood model, the utilization of cellular automata proved helpful. ([23], pp. 148sqq.)

As to the second, Schelling invented a formalized model of the uniform multi-person prisoner's dilemma (MPD) and demonstrated the following result: There is for a given population, some number $k > 1$ such that if individuals numbering $\geq k$ choose their unpreferred alternative and the rest do not, those who do are better off than if they had all chosen their preferred alternatives. This proposition secures a result on the minimum size of successful coalitions of size k. ([23], p.218) For our case of urban behaviour, this proposition is important with respect to the „competitive" co-operation within a certain district community, especially, if organized under the rule of decentralized administration. Obviously, a coalition in this sense, is a subset of the given population that has enough structure to arrive at a collective decision. The problem here is not so much in the inefficient equilibrium of the prisoner's dilemma, but in equilibria which have been achieved by means of unconcerted or undisciplined (i.e. non-communicative) actions. In these cases, everybody would be better off, if there were organized or regulated decision making. (Cf. [23], p.225) Various aspects of this have been discussed in terms of algorithmic decision making or game theory. As it turns out, it is the explicit path dependence of both the information flow and the decision making which introduces hermeneutic rather than merely logical elements into the argument. (Appendix 3)

There is still another, more radical view to all of this: As Sartre has shown some time ago [26], mainly basing his insight on earlier work of Henri Lefebvre [27], [28], the *counterfinality* of individual actions with respect to their collective outcome is one of the most characteristic ingredients of social interactions. In fact, a whole epistemology is being based on this fact of counterfinality dominating most of Sartre's late works. ([26], pp.455sqq., [27], pp.50sq.) What he essen-tially does is to associate the concept of social space with the urban structures generating social behaviour of a given type. Discussing the situation immediately before the outbreak of the French revolution, he introduces the urban location (i.e. a district of Paris, St.-Antoine) as what he calls „practico-inert" tension and *hexis* of the collective of citizens, at the same time. In fact, the collective is being formed into a totality by an exterior organizational praxis which exerts pressure on it (by means of actually drawing troups together between Versailles, where the National Assembly is gathering, and Paris). The point is that while developing the conditions for the critical onset of an avalanche of violence eventually triggering, among other things, the storming of the Bastille, Sartre can demonstrate how it comes that all individual persons participating in the actions which are producing this result in the end (which we nowadays refer to as a prominent „historical event"), in following their own multifarious interests, goals, and objectives, do *not at all* intend any of what will become real in the outcome, hardly knowing in fact what happens, not to speak of any definite political perspective with a long-term range of actions. ([26], pp.449-746) The idea is that the common history of an escalating succession of events is what transforms the original seriality of the citizens into a group in fusion, and this latter *is in fact the city*. This is mainly so because the mentioned succession of events outlines a specific *hexis* as representing a distinctly hodological determination of the socially lived space of the respective quarter which might eventually turn into an explosive *praxis*. ([26], p.462)

It is not the appropriate place here to discuss Sartre's theory in detail. This has been done several times at other places. [29] But what is important for us here is to notice that when asking for the origin and/or explicit dynamics of counterfinality we are led to visualize the very process of communication as the practical means of „averaging out" all these divergent components of individual action so as to eventually letting emerge a collective result which can be viewed as the global outcome of the superposition of many local interactions. And this is very much on the line of Schelling's argument as discussed above. Moreover, as has been shown recently, it appears to be very likely that this „production quality" of communication is in fact a property of language itself which closely binds the explicit linguistic system to characteristic modes of thinking and acting. [30] With respect to Sartre's method this has been discussed in terms of mathematical logic in order to demonstrate where exactly the onset of hermeneutic can be actually situated. In other words: It is this dynamics of counterfinality which indeed produces and forms hermeneutic as chief instrument (and behavioural attitude) in order to achieve a reasonable orientation within a world with incomplete information. [31] This is indeed comparable to what in science is referred to as „decoherence", meaning a process of „averaging out" path variations in a space which consists of all possible transition paths among states of a system. In fact, we can also visualize the social system in terms of such a path space of linguistic type. In this case, we can think of a state as being given by a suitable proposition, the respective paths connecting possible translations of propositions. Note that this is not necessarily a translation among distinct languages. Thinking of interpretation within a social system as constituting the hermeneutically acquired basis for any action, even within one single language, each interpretation shows up as a translation of propositions in this sense. Hence, the social space is loaded with meaning which on the macro-level of a given system shows up as the decoherence of all individual meanings on the micro-level participating in one process in question.

## 4  Urban Structures in Social Space

It is in fact Henri Lefebvre again who for the first time evaluates the social nature of space. [8] The important point is that he relates aspects of social space to the physical nature of the world. In particular, he states that although there is no (apparent) isomorphism between social and physical energies or fields of forces, „all the same, human societies, like living organisms human or extra-human cannot be conceived of independently of the Universe ...“ Hence, „[w]hat can be said without further ado is that the concepts of production and of the act of producing do have certain abstract universality.“ ([8], pp.13sq., 15) This embedding of social systems into physical nature in turn laying the grounds for a self-reference loop of the latter has been recently discussed elsewhere with a general view to recent results in the physical sciences. [32] Important for Lefevbre's interpretation is the aspect of social space turning out to be a social product. And he points out that this fact is however concealed, mainly due to double illusion, namely first, the illusion of transparency, and second, the realistic illusion. The former establishes the fact that communication brings the non-communicated (or: non-communicable) into the realm of the communicated. The latter serves as a re-insuring pleasure which also science can in no wise counteract insofar it does not guarantee the delectable. ([8], pp.27sqq.) One implication of this is that physical (natural) space is disappearing. Nature nevertheless, remains the raw material out of which the productive forces of a society have forged their specific spaces. Another implications is that every society (and thus every mode of production) produces its own space. Hence, social space contains and assigns approrpiate places to the social relations of reproductions and productions. Lefebvre introduces what he calls the *conceptual triad* which consists of spatial practice, representations of space, and representational spaces, respectively. In this sense, abstract conceptual space „presupposes the existence of a spatial economy closely allied to, though not identical with, the verbal (concrete) economy. This valorizes certain relationships among people and places [locations] and thus gives rise to connotative discourses which are able to generate conventions according to which people actually behave.“ ([8], paraphrasing pp.30sqq., 56) These relationships determine a language which models them and carries as its nucleus a logic which is basically one of metaphorization. In principle, the words of any language are simply metaphors for things. (Lefebvre refers expli-citly to spoken language here, very much in the linguistic tradition of Saussure. We can add here with Stetter [30] that this is also true for written language. In fact, the more so, insofar it is the grammar of written language which acts as a logic.) In this sense, „[a s]ociety is a space and an architecture of concepts, forms, and laws whose abstract truth is imposed on the reality of the senses, of bodies, of whishes and desires." ([8], p.139)

In their abstracting from this intrinsic metaphorization of language, objects (products) that are measured can be said to actually lie. They do not speak the truth about themselves, because they are reduced to the common measure of money: „Things lie, and when, having become commodities, they lie in order to conceal their origin, namely social labour, they tend to set themselves as absolutes. Products and the circuits they establish [in space] are fetishized and so become more real than reality itself ... This tendency achieves its ultimate expression ... in the world market." ([8], p.81) Concluding, Lefebvre summarizes that therefore, social space has a part to play among the forces of production, is sometimes consumed and/or productively consumed, is politically instrumental, underpins the relations of reproduction and production and the property relations, is equivalent to a set of institutional and hence ideological superstructures that are not being presented for what they actually are, contains potentialities. ([8], pp.348sqq.)

For the case of architecture, these aspects have been discussed more recently (however without mentioning Lefebvre) by Hillier and Hanson. [33] The idea is to introduce combinatorial measures of urban space based on the assumption of a fundamental „intelligibility of space" which might be drawn from generic shapes empirically observable within this urban space. A semiotic formalization of space is put forward therefore, utilizing an ideographic language which is called henceforth *morphic*. If the natural language has strongly individuated primary morphic units, but a permissive formal structure (syntax), and the formal (mathematical) language very small lexicons instead, and a large syntax (such that it is virtually useless for representing the world as it appears), then the morphic language takes from the formal language the small lexicon, the primacy of syntax over semantic representation, from the natural language the experience-oriented, rule-governed creativity. ([33], paraphrasing pp.48sqq.) The starting point is then one of simulating the onset of clustering of buildings within computer-aided systems of the cellular automaton type due to a small set of simple rules according to the dynamics of a random generator. (Appendix 4) Hillier and Hanson can show then that this simple dynamics can actually reproduce empirically observed settlement structures together with their intrinsic logic. In a new book by Hillier these aspects are formalized in more detail, giving a list of characteristic (combinatorial) parameters for estimating the shapes of settlement structures (and urban spaces, in particular). [34]

The idea is to classify types of buildings and types of their configurations (settlement structures) according to social characteristics which are being mapped in the appropriate shape space. The approach is basically similar to methods of combinatorial topology searching for invariants of shapes and spatial relations among them. It is assumed that there is a set of observable *phenotypes* of structures together with their underlying *genotypes*. Obviously, the genotypes are being thought of referring to characteristic social structures which they are essentially mapping in concrete material terms. We could alternatively formulate that architecture is to buildings what „langue" (the language system) in linguistics is to „parole" (the manner of speaking). In fact, if thinking of the interpretation given by Lacan to signification in linguistic terms, we note a decisive parallel here. In a sense, the mapping relation between the level of social organization and the level of corporal (material) objects is parallel to the respective relation in the theories of Lévi-Strauss between the former level and that of marriage systems, lineages and so forth. Similar to the method of *field re-construction* as applied by Lévi-Strauss when discussing observed social collectives in practice (bricolage), the combinatorial measures as expressed in terms of suitable parameters, easily observable within urban spaces, shall offer an inventory of „first estimates" as starting points for such a re-construction. In the following, we give (according to [34]) a short list of significant parameters, after stating the overall motivation of this:

*Motivation*. Urban layouts are being visualized in terms of more or less deformed grid shapes. The deformation from a regular grid can show up as either *axial deformation* (in that lines of sight and access are blocked by surfaces of building complexes) or as *convex deformation* (in that surfaces vary in their shapes all the time creating a number of patterns). Obviously, the intelligibility of a space will be related to the actual changes in the visibility field of a co-moving observer. This field in turn is determined by convex and axial aspects. The stronger the former, the larger the field, the stronger the latter, the smaller the field. In fact, there will be some kind of locally organizing centre of usually overlapping convex spaces which are called the *integration core* of the settlement. Now, we have then the following list of suitable measures:

*convex articulation* (of space) := number of convex spaces/number of buildings
The lower the values the more synchrony of the space is being achieved.

*grid convexity* := $(\sqrt{I} + 1)^2/C$, where I is the number of islands (blocks of continuously connected buildings), and C is the number of convex spaces.
The lower the values the stronger the deformation from a regular grid pattern.

*axial articulation* := number of axial lines/number of buildings
The lower the values the higher the degree of axiality.

*axial integration* (of convex spaces) := number of axial lines/number of convex spaces
The lower the values the higher the degree of axial integration.

*grid axiality* := $(2\sqrt{I} + 2)/L$, where L is the number of axial lines.
The lower the values the higher the degree of axial deformation.

*convex ringiness* := $I/(2C - 5)$
This measures the number of loops within urban space as compared to the maximum number of possible loops. Insofar it also measures the distributedness of the y-system which can be represented in terms of what is called y-map. This is a map in which convex spaces are being mapped as small circles, together with lines which signify their permeable adjacencies. Practically, a y-map transforms a convex map into a graph. The convex map is the least set of fattest spaces that covers the system. In fact, these are the instruments of visualizing the representation of a space as a set of syntactic relations, both of buildings and of other spaces. The synchrony of a space is then the quantity of space invested in these relations.

*axial ringiness* := $(2L - 5)/I$
This is the equivalent of the aforementioned with respect to axiality.

*axial connectivity* := number of lines a given line intersects
This is a self-explanatory measure of the connected integration of axial lines.

*ring connectivity* := number of rings a given line forms part of
This is the equivalent of the aforementioned with respect to loops.

From these combinatorial measures it is possible to derive a number of map variants explicating various details of the urban space:

*permeability map*
This is essentially a combination of convex spaces with buildings or bounded spaces, together with their connecting lines in terms of adjacency and direct permeability. Obviously, this has decisive consequences for the flow of communication in the urban space.

*decomposition map*
This is a variant of the aforementioned, insofar lines are indicated which link convex spaces constituted by front doors.

To both of the last two maps, converse maps can be drawn to illustrate the absence of flow lines of communication, i.e. locations of isolation.

*justified map*
This map gives the depth of a structure by connecting hierarchically accessible points to a chosen base point. Notably, it is important to analyze the clustering of spaces of equal depth, and the graph patterns of connecting lines.

*relative asymmetry* := $2(MD - 1)/(k - 2)$, where MD is the mean depth, and k the number of spaces considered.

Note that most of these measures can be equally applied to the interior of buildings as well as to exterior spaces. In other words: The structural differentiation of spaces is something which can be continued into the buildings, because their formal partition into (closed) sub-spaces follows the same spatial logic as can be observed within relationships among buildings actually constituting (exterior) urban space. Hence, buildings show up as differentiations of differentiations. This view, namely that urban space can be visualized as the continuation of housing space into the exterior – or viceversa: that buildings and living space within them can be visualized as continuation of urban space into the interior, is an ancient anthropological rule. The same is true for the mapping of explicit social relations: i.e. exterior behaviour maps to exterior structures as interior behaviour is to interior structures. In particular, family systems, socialization rules, segregation procedures, and ritualization, all of them have their equivalents in the formal mapping of (exterior) material objects constituting space.

## 5   Ethical Consequences

From what we have discussed so far, it is quite trivial to immediately trace the ethical implications of the systematic and methodological appoach put forward here. And as we have seen earlier, the actual propagation of one or the other planning procedures are presented to the public with a recourse to ethical aspects of social community life, even with reference to human fundamental rights. But there is another, still more fundamental aspect to this. The point is that the decision for utilizing scientific methods is from the beginning on a decision for choosing an *onto-epistemic* approach to the world. Hence, it is also a decision in favour of acknowledging a rational nucleus of all what humans can perform within their environment. This aspect is important when thinking of the fact that contrary to what is usually claimed, political decisions are almost always based on emotional and explicitly irrational rather than on rational aspects. The reason for this can be seen in what we have discussed referring to Schelling's book. Nevertheless, ethics is the „science" (if you like) of what is adequate. This is the difference between ethics and morality: The former tells us what is adequate or not according to what we presently know, the latter values what is good and bad according to what we presently believe. Hence, politics, or rather its practical application in daily life in terms of elections, planning, executive actions and so forth (we are only talking about political systems which are based on principles of European democratic, legal, and republican thought), is usually judged following the prejudiced structure of traditional morality, and not, as it should be, the objective structure of rational ethics. The point is that the ancient problem of the Big Revolutions of the 18$^{th}$ century (the American, and the French), namely the transformation of the *bourgois* into the *citoyen* has not been solved lately. But if this cannot be achieved, it could be possible perhaps to approach a more modest objective: namely to transform the respective institutions instead, especially, if they are already of the democratic, legal, and republican type. In fact, the idea of the Bologna city government in the sixties was to approach the political pro-blems very much on this line of interpretation. But beyond this layer of everyday politics, there is another, deeper layer of the intrinsic mediation of the ontological and epistemological levels of modelling the world and pratically behaving accordingly. In fact, if the Santa Fe school, or some of its individual protagonists, or protagonists from its vicinity, like Stuart Kauffman, Per Bak, and also Lee Smolin, talk about evolutionary principles on a fundamental level of nature (of which humans are a product), then what they actually do is to aim at this intrinsic mediation and to connect it to an explictly ethical perspective (though the outcome might be different as can be seen by the detailed views of all of them). Hence, to visualize, as we put forward here, the city as an emergent computational system consisting of massively parallel microworlds eventually creating an observable macroworld, means to apply the computational paradigm in an ethical perspective: If there is a computational paradigm, then there is a realistic possibility for an algorithmic approach to the world. It is mathematical logic in fact which provides this algorithm. And although we have to accept that many aspects of human life within social systems cannot be completely modelled according to mathematical procedures what we can do is to notify the logical nucleus in all what there is, and this is the rational nucleus, at the same time.

## 6   The Programme

The starting point for a first taking in sight of the urban phenomenology, can be readily seen as a first trial in qualitatively screening parameters of combinatorial topology as discussed in the sense of Hillier: We choose for this approach the presently available city map which presents more or less the area originally subjected to the sanatation plan of the (nineteen) sixties (Piano per il centro storico, 21$^{st}$ July 1969; 13 sanatation zones of 1970). For a general orientation we give four fragments of the available map scheme signifying grid cells 13 through 16 as those which belong to the ancient city centre. [37] In the following we display these four fragments. They have to be adjoined in a pairwise fashion – each first one of the pair being the Western fragment. (Printing them out, cutting the edges, and re-combining them gives in fact a much better impression of their geometrical shape.) (Fig.1 through Fig.4)

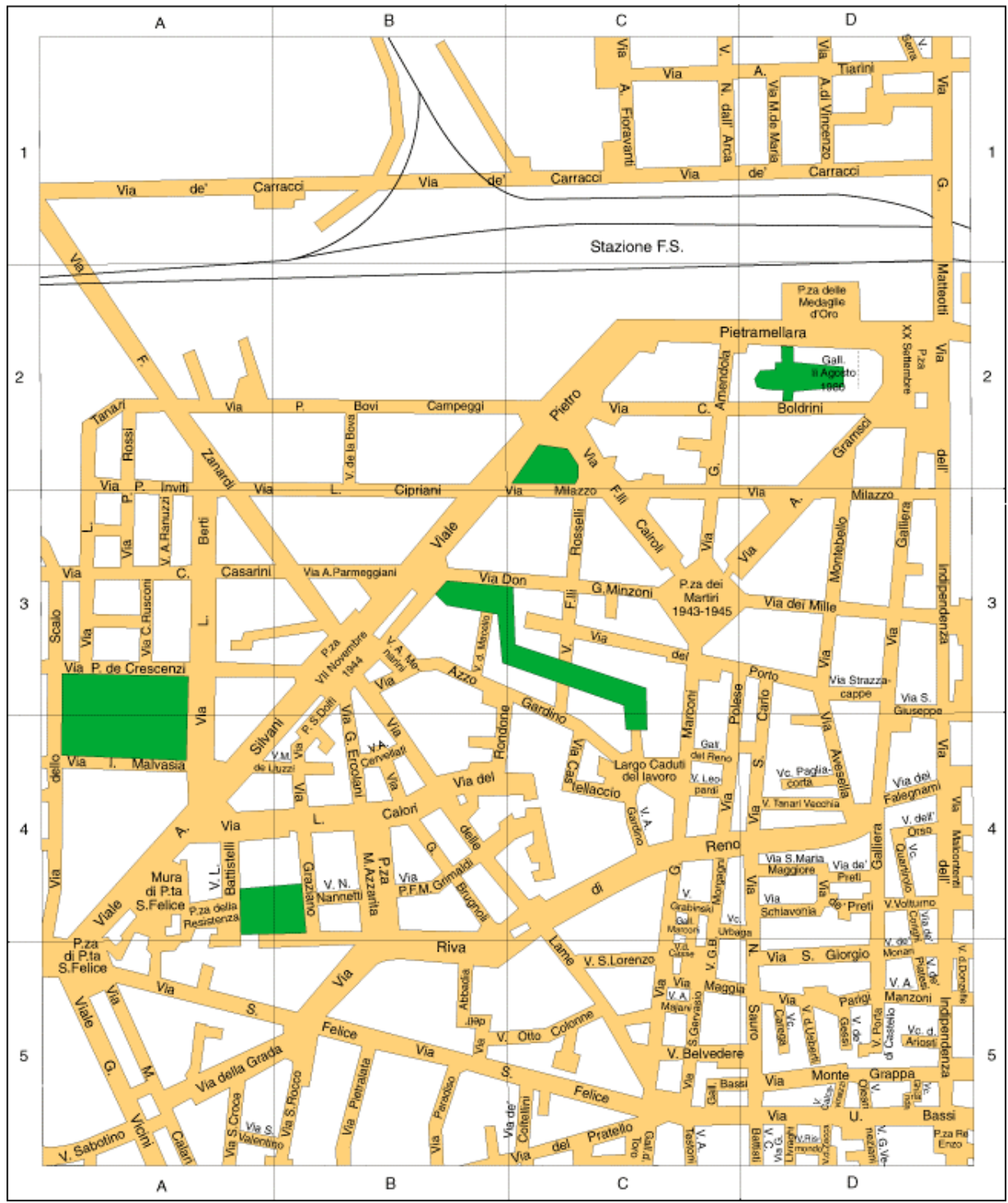

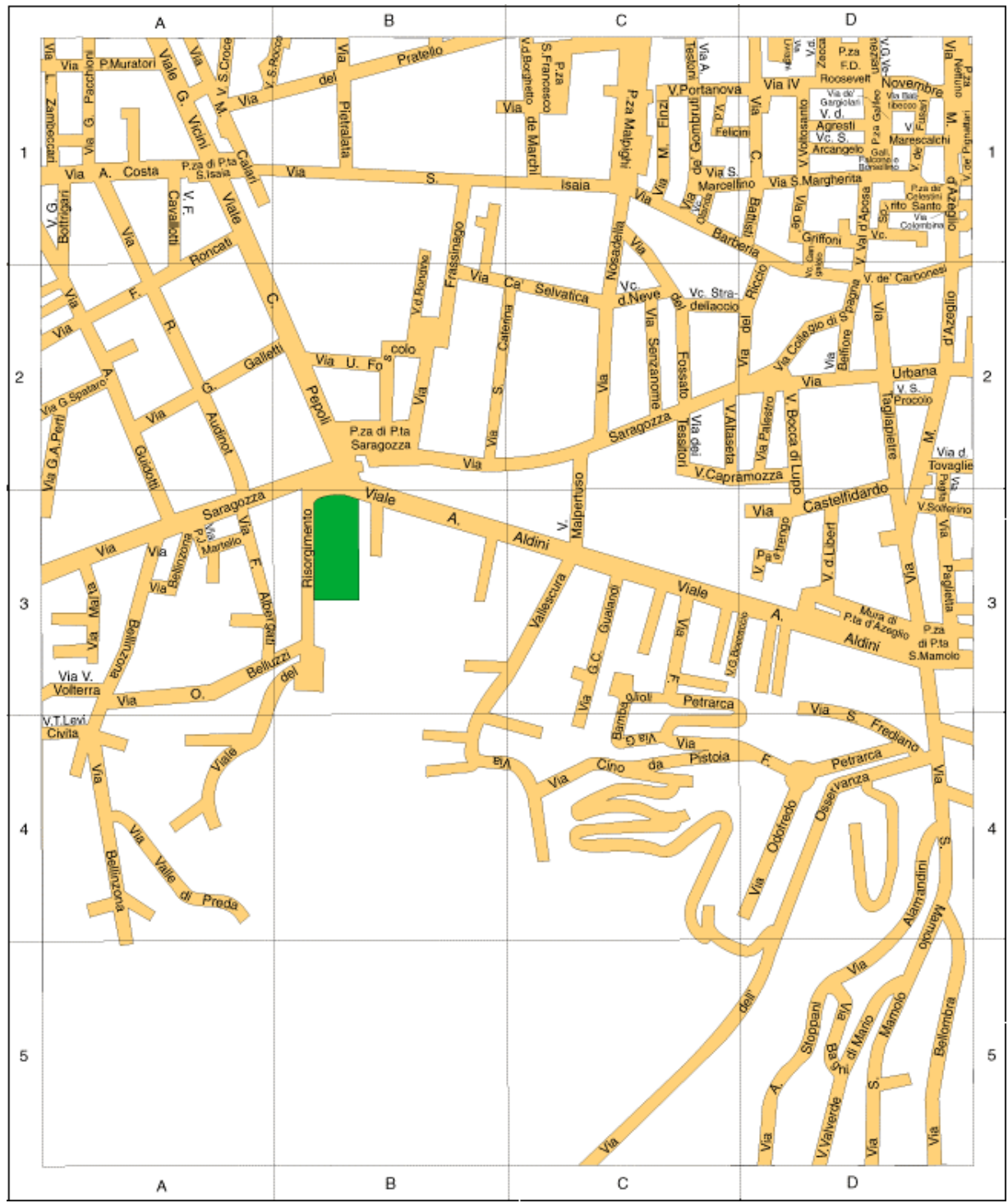

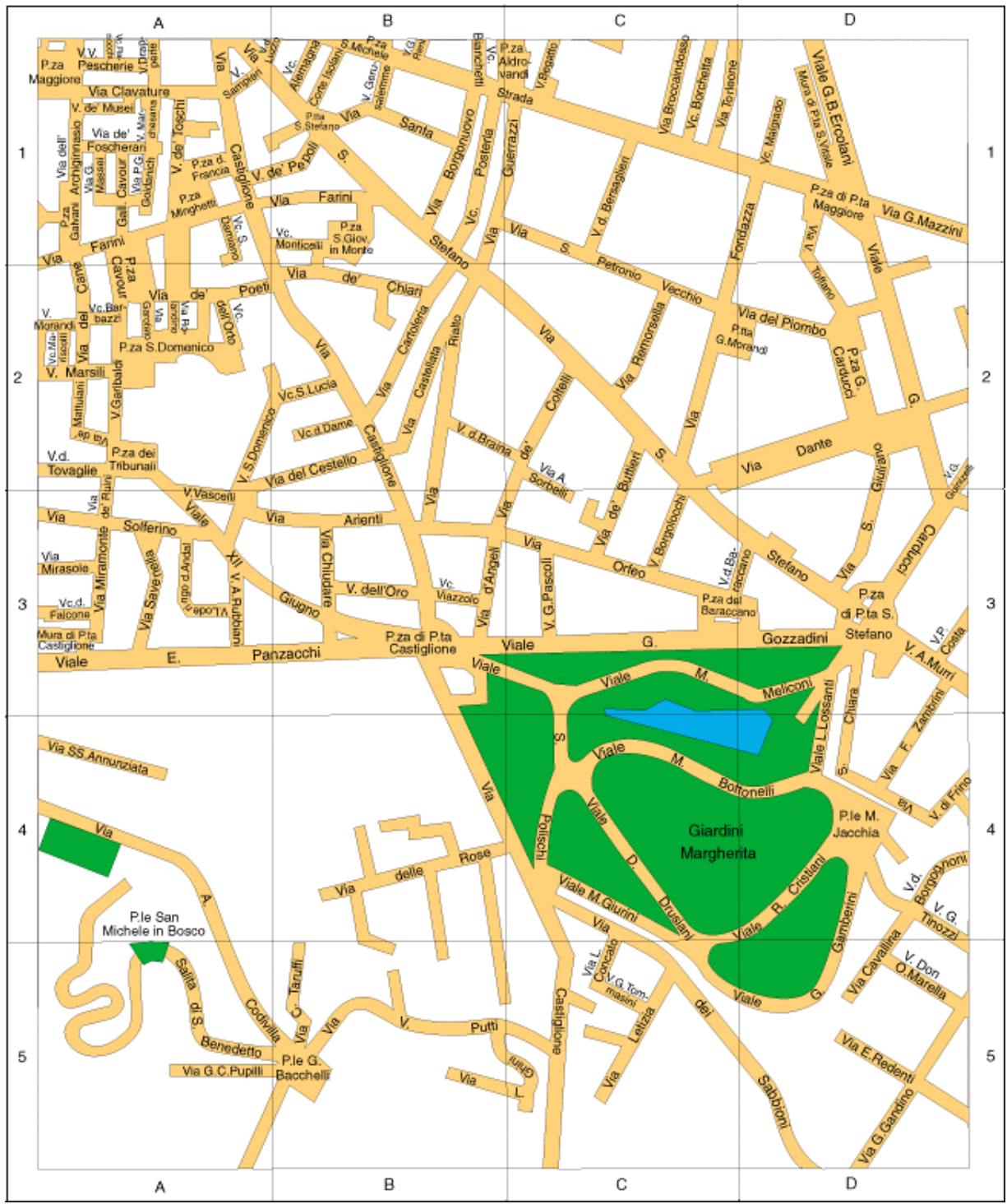

A first look unveals the following interesting points: One can clearly recognize the dominant (not quite spherical) ring structure of the city, essentially reproducing the ancient ring walls. This is the reason for the place names referring to various former gates ( = Piazza di Porta ...). Note in particular, that the southeastern main axis (Strada Maggiore) is identical with the ancient *Via Aemiliana*, one of the Roman Empire's long distance main roads. One can also identify a number of inner ring structures corresponding in a concentric manner to the outset of the outer ring. Choose then the interior one characterized by an obvious increase of street density. What we can find is a not very large convex articulati-on of the space pointing to a relatively high synchrony. The grid connectivity appears to be relatively high which means a low degree of grid deformation. At the same time, it appears that a low break-up is pertinent being characterized by a low value of axial articulation implying a high degree of axiality. In fact, a likewise high degree of axial integration points to a regular grid-like pattern of the urban layout. High convex ringiness is being displayed as compared to medium values of grid axiality and axial ringiness.

A suitable approach to actually interpreting these first intuitive impressions (rather than numerical results which could only be effectively produced by means of computer-aided calculations) is think in terms of „natural movements" in the sense of Hillier. ([34], pp.161sqq.) These movements give simply that proportion of movement on each line that is determined by the structure of the urban grid itself rather than by the presence of specific attractors. In the case of the ancient city of London (which has actually a quite different shape as compared to the ancient centre of Bologna) it can be shown then that despite a high labyrinthian degree of structural organization, the available space does look highly intelligible. This is mainly so, because a large break-up due to many convex spaces is being accompanied by a large number of lines connecting these convex spaces. In other words: Because pedestrians rely on the understanding of the logic of lines they are actually utilizing, the intelligibility of space is being preserved provided the main grid structure is clearly *outlined*. As Hillier formulates: „This means that wherever you go, there is usually a point from which you can see where you have come from and where your next point of aim might be. This is the opposite of labyrinthian." ([34], p.158) This is similar to what one encounters when walking through the streets of Bologna, unlike however, the impression experienced when walking through the streets of Venice.

There is another point to this. Different from other city structures, Bologna is the city of *portici* meaning that there are many arcades skirting the streets adding to the „closed" framed space impression. There are around 37 kms. of *portici* in fact, especially framing the ancient parts of the centre. Obviously, their optical impression created in terms of convexity and axiality is closely related to their originally economic functions which in turn mirror the various social structures. We would assume that this additional *morphological layout* of the urban structure adds to the mediation of public space and domestic (private) space. In fact, the aforementioned aspect of continuing the exterior space into the interior living space can be reproduced in this case when thinking of the explicit courtyard pattern immanent in the ancient centre of Bologna, e.g. in the area of the Strada Maggiore.

This first survey will be the starting point for a more detailed analysis of the interactions of morphological and combinatorial urban structure on the one hand and flow patterns of communication, transport, and economic functions, on the other hand. Continuous modelling might serve here as a first entry. But when dealing with the explicit data available whose discussion will be the topic of the second part of this present series of papers, a second entry can only be provided by means of computer-aided modelling in the sense of the appendix 4.

# Acknowledgements

One of us [R.E.Z] would like to thank John Baez, Louis Kauffman, and Lee Smolin, for stimulating and illuminating discussions during their stay in Cambridge. He also thanks Julian Barbour, Richard Bell, Mary Hesse, Basil Hiley, Chris Isham, and John Spudich for their kind interest and valuable remarks. For co-operation he thanks Tarja Kallio-Tamminen, David Robson, Wolfgang Hofkirchner, and Paola Zizzi, the latter in particular for an invitation to the University of Padova.

# Appendices

## Appendix 1: Negator Algebra & Continuous Modelling

As has been indicated in terms of a somewhat more general perspective in [3] and [4], the urban development can also be visualized as a recursive self-composing process in the sense of negator algebra. As to the relevant variables of quasi-continuous type then, it has been shown in earlier work that (particle number) densities of sub-populations ordered pairwise with their respective „catalyst" group and referring to an information flow model similar to chemotaxis can serve as significant modelling quantities of appropriate scenarios exhibiting the phenomenology of a spontaneous onset of the creation of structure. [11] In the Paris scenario discussed at the time, the idea was to describe the spontaneous creation of a new pair of sub-population/catalyst group (flaneur/bohème) as outcome of a transition of the original („pre-Balzac") scenario from a system of evolution equations expressed in terms of a 7x7 characteristic matrix to a new system expressed in terms of a 10x10 characteristic matrix. Hence, instability of the original scenario and its settling down again into another stable state (a typical sandwich structure of negator type) could be visualized as a kind of mutation in terms of population genetics. This „mutation" could be interpreted then as one which was being triggered by the altered perception of the inhabitants of an urban system undergoing such a transition from „town" to „city", thus a change of social behaviour being accompanied by a change of urban structure. This was technically represented by introducing combinatorial measures of the city structure in purely topological terms: by means of *space contrast* $C = ℏ^2/l^2$, where $ℏ$ is the average height of buildings in the observer's vicinity, and $l$ a typical mean free path of a co-moving observer giving a measure of the relative pedestrian's mobility – also by means of the *labyrinthine coefficient* (with a reminiscence to Walter Benjamin) $κ = c + ½ j$, $c$ being the number of street crossings, and $j$ the number of junctions, respectively. Then, in the evolution equations, the time operator would have to be replaced by the full derivative including the „convective" term taking account of the fact that everything within the city is being observed by an observer who is actually part of the system: hence, $d/dt → ∂/∂t + v \cdot \nabla$, with the „convective" velocity being replaced by the perceived relative average velocity of the form $<u> = κ\, C\, <v>$. Note that this convective term takes the role of a covariant derivative which introduces a geometrical meaning into the time derivative (or change) being coupled to the perception of a co-moving observer. Hence, the concept of *meaning of a social space* is being introduced here explicitly into the operator representation. This is so because the meaning the observer will associate with a given space according to what he perceives will determine the characteristics of his social behaviour. It is the *condensation of city space* therefore, which generically introduces a hermeneutic aspect into the discussion going far beyond the classical analysis of city models. Hence, the importance of preferences when discussing the social relevance of urban planning. This has been recognized for the first time as early as 1977 in the famous urban modelling study of the Prigogine school. [12] This model even incorporated the visual landscape preference of white collar and blue collar workers with a view to their respective choice of housing. Thinking of the various social, political and ethical implications of urban structures in the sense of what has been said in the introduction, no less general (and truly interdisciplinary) model can hope to achieve any relevant insight into the dynamics of urban reconstruction.

Obviously, however, the limit of continuous modelling is in its technical tractability: As it turns out, in the given example things would become more complicated due to the larger number of relevant variables. In the Paris scenario, characteristic matrices of size 7x7 or 10x10 were just at the edge of being treated with respect to deciding about the actual onset of instability (the initiation of the required sandwich structure). In the case of Bologna now, an explicit stratification of social groups would have to be introduced referring to the (original) decentralization in terms of quarters representing clusters of interaction complexes parallel to the pairwise ordering of a group with its associated catalyst. Depending on the demographic structure of, say, the historical centre, this would mean to introduce at least seven social groups (white collar workers, blue collar workers, craftmen, cooperative businesspersons and/or dealers, private ownership (family) businesspersons, public service employees, old age pensioners) stratified according to their living quarters (in this case of four of them: Galvani, Irnerio, Malpighi, Marconi) and associated with their respective catalyst groups (of which intellectuals, artists, journalists, and even vagabonds and homeless might form extra groups, not (yet) so perhaps tourists for that case). In turn, the quarters would have to be classified according to their interaction with their „nearest neighbours" of other quarters (which are Saffi, Bolognina, San Donato, San Vitale, Murri, Colli, and Costa Saragozza). Together with an equation for the information flow, this alone would add up to at least 14 equations for densities and 18 equations for interaction complexes, 33 equations altogether, resulting in a characteristic matrix of appropriate (33x33) size. Unfortunately, this time we would have to add further equations displaying the interactions with the actual traffic flow.

A somewhat simplified, if not naive, model of moderate highway and urban traffic has been given at another place some time ago. [13] This was thought of at the time as a possible variant of a model presented by Prigogine and Herman earlier. [14] The idea was to model the traffic flow according to flow principles of the Navier-Stokes type. This gives for the one-dimensional case of lane traffic the simple equations of the form

$$\rho \, du/dt = - \partial p/\partial x; \quad d\rho/dt = - \rho \, \partial u/\partial x; \quad p = A \, \rho^{\gamma}.$$

Here ρ is the density of the traffic flow ( = particle number density of cars on road), p is the intrinsic pressure satisfying an appropriate equation of state, x is the „length" coordinate dimension, γ the critical exponent of the state equation, and A a constant. The mobility of the flow is determined by the Reynolds number $Re = LU/\mu$, where L and U are characteristic length scales and velocities, respectively, and μ is the viscosity of the flow. Define the Mach number of the flow as u/c with c being the flow's intrinsic „speed of sound" (which is actually identical with the average velocity of the population). What we get in the end, is A ≈ 0.03 for urban traffic, and A ≈ 1.33 for highway traffic. Then c ≈ 30 kmh$^{-1}$ for urban traffic, and c ≈ 100 kmh$^{-1}$ for highway traffic.

The crucial point here is twofold: On the one hand, this simple flow model (whose main interest was solely in stability questions) should be completed in terms of interactions coupling it to the system of population equations as discussed above. On the other hand, for urban traffic, another interactive coupling is even more important: namely that in terms of pollution, noise level, visual intrusion, secondary effects of congestion, particularly for hours of peak traffic, safety requirements, maintenance, and so forth. Obviously, these parameters couple in turn to the aforementioned problems of preferences as to the actual choice of living space within the city centre, land use planning, rentals, sanitation and so on. It appears to be very likely therefore, that a continuous approach to modelling the dynamics of urban systems can at most be useful for collecting first impressions as to the participators in actual interactions. But a computer-aided approach appears to be more effective in the long run, especially when collecting real-life empirical data.

## Appendix 2: Self-Organized Criticality

The basic idea of the theory of self-organized criticality is „that complex behaviour in nature reflects the tendency of large systems with many components to evolve into a poised, „critical" state, where minor disturbances may lead to events, called avalanches, of all sizes." ([19], p.1) The evolution of this state occurs without any design of some kind, it is being established solely because of the dynamical interactions among individual elements of the system. Hence, self-organized criticality is visualized as the generic mechanism of producing complexity. It is a primarily empirical concept. This means that empirical evidence can be collected in order to conclude that a process be of the self-organized critical type or not. In this sense, a phenomenon is said to be self-organized critical, if it exhibits a simple power law structure of its numerical item relationship, of the form

$$N(s) = s^{-\tau},$$

where N(s) is the number of items which have quantity s, and $\tau$ is some power. Taking logarithms then, gives

$$\log N(s) = -\tau \log s,$$

and hence, the logarithms plotted in a diagram give a straight line. This scale invariance turns out to be a characteristic for self-organized critical phenomena. (Cf. [19], p.27)
But there is a number of very fundamental aspects to this concept: The idea of discussing systems of this type has been introduced chiefly by Stuart Kauffman [20] who started to explore whether there might be some general laws governing non-equilibrium systems which could be representing a class of co-evolutionary self-constructing communities of autonomous agents. He argued that evolution could be visualized then as a motion in an appropriate fitness landscape such that the entire system would achieve such a self-organized critical state. As if by an invisible hand, the fitness of each agent would appear then as being maximized. What he did invoke to this purpose was what he called the „fourth law" of thermodynamics: Let us define the set of states that are one reaction step away from those that do already exist as the *adjacent possible*. Then the flow from the actual to the adjacent possible is governed by the „fourth law" stating that the flow is such that the dimensionality of the adjacent possible, on average, expands as rapidly as it can. This would probably conform with a suitable definition of a „time's arrow" in the sense of thermodynamics. On the background of this evolutionary principle then, agents would persistently co-create the worlds they inhabit bringing forth ever changing webs of agents and their niches. (Cf. [20], p.4 from the preface) This also suggests that each interaction space (as is the case with the terrestric biopshere or with the Universe altogether) turns out to be non-ergodic. Then self-organization will take place with respect to three phase transitions: characterizing the „edge of chaos" among agents of a given community, the self-organized critical state among these agents visualized as co-evolutionary system, the (subcritical/supracritical) boundary at which the functional diversity of the community expands, on average, as fast as it can. As Kauffman points out: „The hope is to represent entropy as the information one agent can have about each of its neighbours or its environment. Since the agents are ... co-evolving, this hope suggests a formulation for which entropy is not just a measure of ignorance ... but reflects the shared know-how enabling the system to co-evolve and literally construct co-ordinated properties. / It is at least suggestive that agents share information over their boundaries, hence there are parallels to the fact that the entropy of a Black Hole is proportional to its surface area, and to ideas about entropy being related to the surfaces between different volumes of space." ([20], p.7) These aspects will also be important for our undertaking here, and we will come back to them in due time. Note only that Kauffman's motivation for them can be found in Smolin's book on the „Life of the Cosmos" where Smolin argues that among the decohering quantum histories of the Universe, those in which the Adjacent Possible expands fastest will decohere the most readily. Thus, the Universe might tend to flow towards maximum complexity. As far as it goes, this idea deals with a persistent competition among all possible laws which might govern the Universe such that this competition can be modelled in terms of transformation rules of *spin networks* which serve as the most fundamental quantities constituting the Universe. [21]

It is not the appropriate place here to discuss further details of this cosmological approach. This has been done at other places. [22] But for the study pursued in this present work, it is relevant to note the generality of the underlying laws and the consequences of this: As has been mentioned earlier, it appears to be very likely that global „gross" qualities of dynamical laws in nature are reflected in each specific detail governing local interactions in various fields of science. If we indeed visualize the evolution of a city as the structural unfolding of a self-organized critical and computational system, then obviously, the explicit results gain a decisively pointed obligation and commitment. The city as a social system shows up as *one natural system among many others*. And this is not only compatible with a modern approach to materialistic philosophy, but also with a modern approach as to the position of humans within a natural world of which they are an interacting part.

## Appendix 3: Categories & Hermeneutic

As has been shown in more detail elsewhere [22], [31], it is *mathematical category theory* which offers a suitable approach to formalizing the transition from logic to hermeneutic demonstrating at the same time that the latter can be visualized as a generalized version of the former, for the case of incomplete information. In this sense, the description of processes can be mapped in terms of propositions formulated in some language. Hence, languages altogether can be thought of as being categories whose objects are propositions formed out of a given lexicology and satisfying rules as laid down by grammar, these rules being the category's morphisms. *Translation* then is a functor between language categories mapping objects to objects (propositions to propositions) and morphisms to morphisms (rules to rules). The idea of this approach is that translations are path-dependent with respect to compositions. This can be shown in that the diagram of the form

$$\begin{array}{ccc} A & \rightarrow & B \\ \downarrow & & \downarrow \\ C & \Leftrightarrow & C \end{array}$$

is not commutative, if A, B, and C mean any languages. (The double arrow indicates the identity mapping.) But it is commutative, if A, B, and C refer to diffe-rent logics. Leaving aside the process of actually producing knowledge (which is not our topic here), we can say that in the case of the logics, the information (with respect to the semantics of the propositions utilized, e.g. with a view to applications in physics) is complete. Hence, translation of propositions is commutative with respect to the diagram shown above, if we deal with logic. It is thus *path-independent*. But for hermeneutic, which shows up here as a generalized logic for incomplete information, it is *path-dependent*. (To turn this the other way round: Logic is a hermeneutic with complete information. Each hermeneutic has thus its logic nucleus: This means nothing is completely arbitrary. There is always a rational aspect which is at the foundation of what is actually being done.)

Define the *action function* of communication in terms of a suitable energy balance of states to be of the form $S = \int L \, dt$, where L is some function defined on the phase space representing the set of all possible states of a given system (e.g. a social group). This could be thought of as a kind of energy freely available to produce meaning out of communication in the translational sense as disucssed before. Then, in principle, for each individual, there is one distinct possibility to translate one proposition into another in order to determine its meaning (i.e. interpreting it). And all individuals participating in the process of communication will perform these interpretations at the same time. The actual result of this process is achieved by „averaging out" all these distinct views by means of *decoherence*. This can be understood in terms of defining a path integral such that

$$\int \exp k \, S \, d\Gamma,$$

where k is a constant, and the integral is taken over all paths in the path space $\Gamma$. The idea of this is that the path integral takes care of all individual interpretations, but that only that one path is actually being observed as the outcome (or consensus or governing opinion) of the communication process for which the variation of the action function vanishes, i.e. $\delta S = 0$. This is obviously an optimality condition which produces the superposition involved and which in the end is nothing but the „historical interpretation" creating a „historical event". All these aspects are particularly relevant as to the application of induced strategies (such as de-centralization) with respect to which communication turns out to be rather sensitive.

## Appendix 4: The StarLogo Approach

There are many computer-aided simulation programs available, sometimes of a rather complex kind. The *Transims* project e.g., of the Los Alamos National Laboratory, is one of the more complex types actually simulating wide range transport processes including pollution, energy consumption, land use planning, and so forth, for metropolitan areas such as that of San Diego. [35] For more limited computational resources it is advisable to utilize more qualitative simulation programs also based on the principles of cellular automata. One such approach has been introduced by Michael Resnick. [36] The important point is that he also starts from fundamental assumptions about evolutionary processes. One basic assumption is that of the paradigmatic value of *de-centralization* which takes a prominent role in our analysis here, as we have shown above. Consequently, Resnick refers to Schelling's book. ([36], pp.9sqq., 86) There is actually the possibility to download the StarLogo software and a catalogue of sample sessions from MIT's medialab web page: http://www.media.mit.edu/. (Note that the address given in the book, p.153: /~starlogo, is apparently not more correct so that the download should be initialized by tracing the true address from the main menu.)

The basic idea of the approach is to deal with three fundamental „characters" or object classes, called turtles, patches, and observers. The turtles are the inhabitants of the StarLogo world, entities capable of executing procedures, interactions, self-reproduction and production. The patches are fragments of this world which can be self-active in also being capable of executing commands. This functions very similar to a cellular automaton so that the StarLogo world can be visualized as an inhabited cellular automaton. The observer is an exterior (godlike) observer of the system being able to create new turtles and to give commands to turtles and patches, and to actually monitor what happens. The set of commands in StarLogo is a collection of usually abbreviated elementary directions. There is also the possibility to run demons which are continuously run background programs serving to establish a process parallelism accompanying the data parallelism given by the main program procedures. Turtles can be differentiated according to their splitting up in various different populations. And a colouring system enables the symbolically loaded colouring of patches.

There is a large collection of sample procedures capable of simulating evolutionary scenarios whose programs can be obtained from the book (and are actually available in the sample catalogue of the download version). This comprises of prominent examples of self-organized criticality including the famous „slime molds", „artificial ants", „traffic jams", and „forest fires" (the original example for a celebrated percolation model).

Note that many of the processes we are actually dealing with here in our specific example can be visualized in terms of flow phenomena or percolation. Transport and information flow are examples for that. Hence, the idea is to start the modelling with simplified StarLogo procedures as a second step following up the discussion of first qualitative measures of urban space, and the utilization of methods stemming from continuous modelling.